\documentclass[preprint,12pt]{elsarticle}

\usepackage[title,toc,titletoc,page]{appendix}
\usepackage{graphicx}
\usepackage[font=small,labelfont=bf, justification=justified,
   format=plain]{caption}
\usepackage{subcaption}
\usepackage{amssymb}
\usepackage{amsmath}
\usepackage[]{algorithm2e}

\DeclareMathOperator*{\argmax}{argmax} 

\makeatletter
\def\ps@pprintTitle{%
 \let\@oddhead\@empty
 \let\@evenhead\@empty
 \def\@oddfoot{}%
 \let\@evenfoot\@oddfoot}
\makeatother

\begin{document}

\begin{frontmatter}


\title{Robust Propensity Score Computation Method based on Machine Learning with Label-corrupted Data}


\author[label1]{Chen Wang}
\author[label2,label3]{Suzhen Wang}
\author[label2]{Fuyan Shi}
\author[label2]{Zaixiang Wang}
\address[label1]{Dept. of Computer Science, University College London, Gower Street, London, England, United Kingdom  WC1E 6BT}
\address[label2]{Dept. of Health Statistics, Weifang Medical College, No.7166 BaoTong West Street, Weifang, Shandong, China 261053}
\fntext[label3]{Corresponding Author}
\cortext[cor1]{This research is granted by China National Natural Science Foundation, Grant: 81473071}
\begin{abstract}
In biostatistics, propensity score is a common approach to analyze the imbalance of covariate and process confounding covariates to eliminate differences between groups. While there are an abundant amount of methods to compute propensity score, a common issue of them is the corrupted labels in the dataset. For example, the data collected from the patients could contain samples that are treated mistakenly, and the computing methods could incorporate them as a misleading information. In this paper, we propose a Machine Learning-based method to handle the problem. Specifically, we utilize the fact that the majority of sample should be labeled with the correct instance and design an approach to first cluster the data with spectral clustering and then sample a new dataset with a distribution processed from the clustering results. The propensity score is computed by Xgboost, and a mathematical justification of our method is provided in this paper. The experimental results illustrate that xgboost propensity scores computing with the data processed by our method could outperform the same method with original data, and the advantages of our method increases as we add some artificial corruptions to the dataset. Meanwhile, the implementation of xgboost to compute propensity score for multiple treatments is also a pioneering work in the area.
\end{abstract}

\begin{keyword}
Biostatistics \sep Propensity Score \sep Spectral Clustering \sep Data Sampling \sep Xgboost \sep Machine Learning


\end{keyword}

\end{frontmatter}


\section{Introduction}
Confounding covariates, or alternatively named as noise features, is a significant problem in studying Biostatistics data. In practice, data in this field is usually collected with detailed information, thus it will contain some irrelevant and redundant features. For example, in a dataset of a certain disease, the information about the marital status might be irrelevant to the cause of the disease, and it will be acting as a confounding (noise) covariate. The confounding covariates could bring negative effects in analyzing the data: it could provide misleading information to reveal a false correlation between variables (features) and labels, and it could introduce unnecessary statistical differences between groups and lead to a incomparable situation. Consequently, recognizing and balancing confounding covariates have become an open task in biostatistics research.\\
A common approach to address this problem is propensity score \cite{RosenbaumRubinOriginal}. The key idea of propensity score is to compute the 'likelihood'(propensity) of a sample to be belonging into a certain group (usually termed as 'treatment' in biostatistics). And by matching the samples in different groups with similar propensity scores and/or weighting the samples with propensity score, we could reduce or eliminate the bias of confounding features. \cite{MengPropensityExperiment} experimentally shows that for some randomly generated data, propensity score could effectively eliminate the statistical significance between groups. There are multiple methods to compute propensity score, ranging from simple logistic regression to cutting-edge Machine Learning methods. From the perspective of Machine Learning, the task of propensity score computing is essentially to train and utilize a classifier with probability as retrievable outputs (like soft-max output). Consequently, although there are a variety of Machine Learning algorithms which could be plunged to the task, the basic procedures are almost the same.\\
While the state-of-the-art Machine Learning methods are promising in this area, they typically suffer from one problem: the corrupted data. Data is typically collected from medical practice, and there could be some mistreated samples in each group. The conventional approach to solving this problem would be to hire some experts to select the correctly-labeled samples and use them as a subset to train the classifier. However, this procedure will be expensive in both time and money. \\
To tackle the issue with a smarter approach,  we propose a novel Machine Learning-based propensity score computation procedure in this paper. We exploit the fact that the majority of samples should be treated (classified) with the correct group (label) and patients (samples) be treated with different groups should have distinct data manifolds. Guided by this premise, we design a system that first clusters the data with spectral clustering method, and then computes the weight of the data by computing the proportion of the number of samples through different classes (treatments). The weights are processed with an interpolation method to perform a 'moderate softmax' computation and it will lead to a distribution over clusters after being normalized. Based on this new distribution, we will sample a subset for the usage of the training procedure. And finally, the propensity score is computed based on Xgboost proposed by Chen and Guestrin \cite{XgBoostPaper}. \\
The rest of the paper is arranged in the following order: the second section will briefly review literatures related to our research; the third part will introduce the proposed system and explain the Machine Learning algorithms applied; the fourth portion will be mathematically analyzing the assumptions and properties of the proposed method; the experimental result for the proposed method based on SEER data is illustrated in the fifth section, and the outcomes are analyzed and compared; and finally, the last section of the paper concludes our research and discusses future research for this topic.
\section{Related Work}
Despite the solid theoretical foundation of conventional Logistic Regression method in computing propensity score, efforts to utilize novel Machine Learning algorithms in computing propensity score has emerged for a long time. Huge research potentials of propensity score could be found in the state-of-the-art algorithms, which could significantly outperform the traditional Logistic Regression. Existed papers related to this topic mainly focus on implementing Machine Learning methods with advanced models (such as SVM and Random Forest) and comparing different approaches with certain datasets. \cite{WESTREICH2010826} tested different Machine Learning methods in computing propensity score and demonstrated that boosting methods and tree algorithms could outperform Logistic Regression in general. \cite{MLBiostaTheory} studied the theoretical foundations of implementing Machine Learning methods for dichotomous and multicategory outcome in biostatistics, and analyzed algorithms like kNN, Random Forest and SVM. \cite{ClassificationDecisionTree} compared the propensity score results of Classification Tree Analysis (CTA) and Logistic Regression and argued that although Logistic Regression method still has the lowest average standardized difference, CTA could provide the greatest predictive accuracy thus it has strong potentials and could provide an alternative way in propensity score computation. And for the tree-based method specifically, \cite{TreeMethodsPropensity} compared different classification and regression and discussed their performances.\\
There are also published literatures concentrating on improving the model complexity to avoid potential model misspecified problem. A typical example of these algorithms is the so-called 'super-leaner' algorithm, which is roughly equivalent to stacking algorithms in Machine Learning. The method was originally proposed by \cite{SuperLearnerProposal} to process data with complex relationship between input and output through a weighed combination of different learners. \cite{SuperLearnerAnalysis} shows experimentally that the super learner method could achieve a significantly better performance under the situation of severe model misspecification. \cite{SuperLearnerCombination} proposed an algorithm to use a convex combination of generalized propensity score computed by Super Learner. The authors then takes traumatic brain injury as a study case and discussed the statistical significance between time for the patient to be transferred from emergency to specialist and the treatment effect. \cite{2017SuperLearnerHealthCare} further studied the Super Learner method in the application of electronic healthcare data.\\
Whilst there are a large amount of work discussing advanced algorithms in computing propensity score, approaches dealing with propensity score under corrupted data is relatively under-developed. \cite{PropensityScoreMissingData} analyzed the problem of propensity score with missing data and suggested to use multiple imputation method. Similar with this paper, \cite{MixtureModelMissingCovariate} analyzed propensity score computation under missing covariate with multiple imputation and specifically solved the issue with general location mixture models. To the best our knowledge, hitherto, there has not been any significant publication in studying propensity score with mislabeled data, although this is common in bioscience. In the area of Machine Learning, this issue could be regarded as corrupted label problem, and there are some related literatures building algorithms for this task \cite{NIPS2013_5073}\cite{KnowledgeDissilationCorruptedLabel}. However, in this work, we do not follow the idea of advanced algorithmic target functions employed in these methods. Instead, we used a clustering and normalization based method to tackle the problem. Based on our assumption, the method should be valid for biostatistic data and the rationality of the method is shown mathematically in our paper.\\
\section{Methodology}
\subsection{Problem Setup}
Assuming we have data $(X,Y)$ that generated from space $\mathcal{X} \times \mathcal{Y}$, where $\mathcal{X}$ is the input space and $\mathcal{Y}$ is a discrete output space. We now have $m$ data pairs $(X_{1},y_{1}),(X_{2},y_{2}),...,(X_{m},y_{m})$ drawn from a certain distribution. Then, there will be an underlying conditional distribution:\\
\begin{equation}
p(\boldsymbol{y}|\boldsymbol{X};\boldsymbol{\theta})
\end{equation}
And to infer the probability of each data point, we need to obtain parameter $\theta$ that determines the model. With proper hypothesis space, the conventional MLE of $\theta$ should be:\\
\begin{equation}
\theta^{*} = \argmax_{\theta} \prod_{i=1}^{m}p(\boldsymbol{y_i}|\boldsymbol{X_i};\boldsymbol{\theta})
\end{equation}
However, here our situation is the set $Y$ has been replaced by $\hat{Y}$, which contains many noisy(incorrect) labels. Then directly learning from $(X_{1},\hat{y_{1}}),$ $(X_{2},\hat{y_{2}}),...,(X_{m},\hat{y_{m}})$ will incorporate some noise. Thus, it demands an approach to solve the problem.\\
Given the research context of medical and biological science, we could assume the probability for a sample to be labeled (treated) correctly is higher than the reverse situation. Mathematically, this assumption could be noted as:\\
\begin{equation}\label{MajorityAssumption}
p(\hat{y}=a|y=a)>p(\hat{y}\neq a|y=a), 	\forall a \in \mathcal{Y}
\end{equation}
And under this assumption, we could try to sample a subset of $(X_{1},\tilde{y_{1}}),$ $(X_{2},\tilde{y_{2}}),$ $...$ $,(X_{k},\tilde{y_{k}}),$ $k\leqslant m$ to fit the model, and find the parameter with:\\
\begin{equation}
\tilde{\theta} = \argmax_{\theta} \prod_{i=1}^{k}p(\boldsymbol{\tilde{y_i}}|\boldsymbol{X_i};\boldsymbol{\theta})
\end{equation}
that could minimize the loss $\mathcal{L}(\tilde{\theta},\theta^{*})$ between the estimated parameter and the  parameter estimated under uncorrupted data. However, since the true data set is not visible for us, it is impossible to directly optimize over $\mathcal{L}(\tilde{\theta},\theta^{*})$ function. An alternative way would be to sample from a distribution that could produce a subset close to the true dataset, and this is the core idea of our noise-eliminating procedure. The next subsection will be introducing this method and the employed Machine Learning algorithm.
\subsection{Avoid Noisy Label: Sampling with Spectral Clustering and Normalization}
Now we need a method to sample a subset of data with the distribution as close as possible to the original correct samples. The method should have the following properties:\\
\begin{itemize}
\item The proportion of mis-specified labels should monotonously decrease as the performance of the algorithm increase
\item The result should not be affected by the proportion of data in each class (treatment)
\item The results of the algorithm should be interpretable, and the interpretation should have a fixed methodology
\end{itemize}
To satisfy the above principles, we could assume some special properties for the data. Without losing generality, we could assume the data, which is classified (treated) with $d$ different classes, comes from $n$ underlying patterns under condition $n\geqslant d$, and each pattern will have a proportion of $p_{i},i=1,2,...,n$ to be classified into class $j\in \{1,2,..,d\}$. Formally, we could specify the number of true samples under each class $j$:\\
\begin{equation}\label{BasicAssumption}
N_{true}(j) = \sum_{i=1}^{n} {p_{i}N(i)}
\end{equation}
Where $N(\cdot)$ denotes the number of samples. In practice, we choose the number of $n$ as $n=d$ for a better interpretability, and this setting will be further discussed in section 5 with the example of SEER data. Under this assumption, we will be able to sample the data with their underlying patterns, which could be represented by clusters under the skeleton of unsupervised learning. Intuitively, an effective method would be to sample from the clustering results, and now the question is which kind of technique should we use for clustering. \\
The commonly-used clustering method is K-means and its variations. This branch of algorithms are straightforward to implement and the computational complexities of them are generally acceptable for ordinary computers. However, K-means related algorithms suffer from several problems: it is sensitive with initial clustering centers, and in practice we usually need to run K-mean for several times with different initial clustering centroids to obtain the best performance; it is vulnerable towards outliers, as few outliers will make significant effects on the result; and most importantly, the clustering results of K-means branch algorithms have a strictly spherical curvature, thus it could only deal with simple data manifold, but in bio-medical science the data manifold could be of great complexity. \\
A better alternative would be Spectral Clustering. Instead of measure the Euclidean distance between samples, which will inevitably loss some manifold information, Spectral Clustering method utilize the Graph Distance that could preserve local neighborhood informations. In Spectral Clustering, the affine matrix that denotes the Graph Distance and its corresponding Graph Laplacian will be computed, and then eigenvalue decomposition will be performed to capture the information in the so-called embedded space and the clustering result could be obtained by applying 'ordinary' clustering methods on this processed matrix. Spectral Clustering works especially well when the number of clusters is not large, and it could deal with complicated data geometries. The core problem of Spectral Clustering is the Neighbor Graph (Similarity Graph), which represents the 'similarity' between different data points. Common methods to be implemented include $\epsilon$-Neighbourhood, $k$-nearest Neighbourhood and fully-connected graph with positive semi-definite kernels. One popular similarity graph matrix construction method is Gaussian Kernel method, which computes the similarity between data points $\boldsymbol{x_{i}}$ and $\boldsymbol{x_{j}}$ with:\\
\begin{equation}\label{GaussianKernel}
K_{g}(\boldsymbol{x_{i}},\boldsymbol{x_{j}}) = \text{exp}(-\frac{||\boldsymbol{x_{i}}-\boldsymbol{x_{j}}||_{2}^{2}}{2\sigma^{2}})
\end{equation}
Which $\sigma$ is a parameter that could be specified or computed by the following equation:
\begin{equation}\label{SigmaComputing}
\sigma = \sqrt{median(\mathbb{I}(||\boldsymbol{\hat{X}}-\boldsymbol{\hat{X}}^{T}||^2))}
\end{equation}
Where $\boldsymbol{\hat{X}}$ is a $m \times m$ matrix with $\boldsymbol{\hat{X}}_{ij} = ||\boldsymbol{x}_{i}||_{2}^{2},\forall j\in {1,2,...,m}$ (duplicated column matrix with each row as the square $l_2$ norm of the data point) and $\mathbb{I}$ stands for the operation to remove the diagonal values of the matrix.\\
The method could be computed with high efficiency for high dimension data. However, it could be of great instability if some of the features are discrete values. To improve the clustering method specifically for our data with a mixture of continuous and discrete features, we proposed a feature-wise Similarity Matrix method here. For each pair of samples $(\boldsymbol{x_{i}},\boldsymbol{x_{j}})$, we compute the sum of the similarity across each feature, which could be denoted by:\\
\begin{equation}\label{GraphDesignEmbed}
K(\boldsymbol{x_{i}},\boldsymbol{x_{j}}) = \sum_{p=1}^{P} K_{d}(x_{ip},x_{jp}) + K_{g}(\boldsymbol{x^{*}_{i}},\boldsymbol{x^{*}_{j}}) 
\end{equation}
Where $P$ is the number of discrete features(co-variates) and $K_{g}(\cdot,\cdot)$ stands for Gaussian Kernel for continuous features computed by equation \ref{GaussianKernel}. For the discrete features, we compute each of the similarity matrix with Delta Kernel:\\
\begin{equation}\label{DeltaKernelCompute}
K_{d}(x_{i},x_{j}) =
\begin{cases}
    \frac{1}{N(x)}, & x_{i} = x_{j}\\
    0,              & \text{otherwise}
\end{cases}
\end{equation}
Where $N(x)$ is the number of samples of the specific class. And since all the components of the similarity is positive semi-definite, the resulting matrix would also be positive semi-definite. The Process of the detailed algorithm could be shown in algorithm \ref{SpectralClusteringAlgorithm}.\\
\begin{algorithm}[t]
 \KwData{Input data, separated by continuous features and discrete features}
 \KwResult{Clustering result for each data point}
 Compute the affine matrix of continuous features based on equation \ref{GaussianKernel} \;
 \For{Each Discrete Feature}{
    Compute the affine matrix of discrete features based on equation \ref{DeltaKernelCompute} and sum them with equation \ref{GraphDesignEmbed} \;
    }
 Perform Spectral Clustering with the affine matrix \;
 Return the clustering label for each data \;
 \caption{Spectral Clustering of the Data}
 \label{SpectralClusteringAlgorithm}
\end{algorithm}
The practical implement of this method will be discussed in section 5. And after clustering, we could further divide each cluster with sample in different classes(treatments) and get a $d\times n$ matrix:\\
\begin{equation}\label{AssignMatrix}
\boldsymbol{W} = 
\begin{bmatrix}
    w_{11}      & w_{12} & \cdots & w_{1n} \\
    w_{21}      & w_{22} & \cdots & w_{2n} \\
    \cdots        \\
    w_{d1}     &  w_{d2} & \cdots  & w_{dn}
\end{bmatrix}
\end{equation}
Where $d$ is the number of classes and $n$ is the number of underlying patterns (clusters).\\
Intuitively,now we would be able to choose the top $k$ clusters at each class(treatment) and use them as the new training set. However, the number of samples in each cluster, which serves are an indication of underlying patterns, could be imbalance and misguiding in our circumstance. A smarter way would be to normalize the quantity of each cluster across the classes, which could be representation as:\\
\begin{equation}
w^{*}_{ij} = \frac{w_{ij}}{\sum_{i=1}^{d}w_{ij}}
\end{equation}
This will give us the exact information about 'how large is the portion of this pattern being assign with treatment $i$', and it could best reflect and satisfy the assumption denoted by equation \ref{BasicAssumption}. And here we need to augment the probability of the cluster with largest probability because we have mis-specified data in each treatment class. Based on the assumption of \ref{MajorityAssumption}, the augmented probabilities should be belonging to the clusters that should contribute to the treatment. And inspired by \cite{AlternativeOfSoftmax}, we use an interpolation method to combine the first and second order value of $w^{*}_{ij}$ to get $\hat{w}$:
\begin{equation}
\hat{w} = w^{*}_{ij}+\frac{\gamma}{1-\eta * \epsilon} {w^{*}_{ij}}^2
\end{equation}
Parameters $\gamma$ denotes the trade-off between first and second order interpolation, and $\epsilon$ is the proportion of corrupted labels (default as 0 even with some mild corruptions, and we assume $\epsilon \leq 0.5$) and $\eta$ is a parameter denoting how will we treat the importance of data corruption. Here we also have a constraint to $\eta$ for $0 \leq \eta \leq 2$. In our program we set $\gamma=0.7$ and $\eta = 2$. And after the normalization, we could sample a new dataset with the proportion of $\hat{w}$ in each row of the matrix. Formally, it could be denoted by $p^{*}_{ij}$:
\begin{equation}\label{FinalSampleEquation}
p^{*}_{ij}  =\frac{\hat{w_{ij}}}{\sum_{j=1}^{n}{\hat{w_{ij}}}}
\end{equation}
Equation \ref{FinalSampleEquation} is the equation we finally used in computing the probability to sample a subset from each treatment class. In section 4 we will discuss the rationality of it mathematically, but in this section, we will continue on discussing the technique to compute propensity score.\\
\subsection{Propensity Score Computing with Xgboost}
As it is stated in section 2, implementing Machine Learning techniques in computing Propensity Score has been studied for a long time. Propensity Score computing is very close to the probabilistic output in supervised classification problem, thus a large amount of logistic or softmax function-based approach could be introduced into the Propensity score area. Previous studies generally suggest that boosting method is one of the most promising branches of ML algorithms in computing propensity score. And in our work, we use Xgboost as a novel Machine Learning method (designed in 2016) to compute propensity score.\\
Xgboost is actually not a newly-proposed algorithm. Instead, it is built on the algorithm of Gradient Boost Machine (GBM), and the novelty of this platform is that it considered Structural Learning scores and designed a better interface for Machine Learning considerations. The platform has R, C++, Python, Scalar and Java API, and it could be conveniently modified by simply change some parameters and arguments. It support softmax output for multiple-class probability, but using softmax could be too extreme. The package also provide another mode called 'softprob' model, which could output the normalized probabilities of each class. And this is also the method we adopted in this work.\\
Key parameters of Xgboost also include $l_2$ regularization parameter, named 'lambda', $l_1$ regularization coefficient, named 'alpha', and tree construction and growth method, which is often selected automatically by heuristic search. Meanwhile, since we are asking the Xgboost to output multi-class result, here we use multi-class log loss to denote our training and validation loss. The loss could be expressed by:\\
\begin{equation}
l(\boldsymbol{Y},\boldsymbol{P}) = \sum_{i=1}^{m}{\boldsymbol{y}_{i}^{T}log(\boldsymbol{p}_i)}
\end{equation}
Where $\boldsymbol{y}_{i}$ is the $n \times 1$ one-hot vector that has value 1 at the true output position and 0 otherwise; $\boldsymbol{p}_i$ is the probability output vector. $\boldsymbol{Y}$ and $\boldsymbol{P}$ are the $m \times n$ matrices that denotes all the data.\\
Xgboost is a well-developed platform which is 'efficient, flexible and portable' (from their Github page). Here, we will not elaborate on the basic principles of the GBM algorithm it utilizes. One who interested in could refer to the original publication of the method.
\subsection{Data Manifold Analysis with t-SNE}
Another point we might interest in is the visualization of the data, which could bring significant benefits for result interpretation. Visualizing the projection of the Manifold of data could provide us information about how data in different classes could be overlapped or separated. And sometimes it could even help us determine whether there exists significant difference between different groups (If there is not the experiment is probably unnecessary). \\
Typical data visualization approaches often project the data manifold to a 2-d plane, which could be straightforwardly analyzed by human. Popular methods used to compute data visualization include PCA(and pPCA, kPCA, etc.), Isomap, Sammon Mapping, SNE(tSNE) and auto-encoder. The amjor drawbacks of PCA and mapping methods are that they could not preserve local structures and will thus lose the information of the data manifold in the higher dimension. On the contrary, SNE(tSNE) could manipulate the graph neibourhood structure. It is similar with Spectral Clustering method we mentioned before, the difference is that here we compute the data point in the lower dimension instead of clustering the Graph Laplacians. \\
To begin with, we will start to introduce SNE method. SNE stands Stochastic Neighbour Embedding, and it start with the transition matrix between different data points in the original space, computed by:\\
\begin{equation}
p_{i|j} =
\begin{cases}
\frac{\text{exp}(-||\boldsymbol(x_i)-\boldsymbol(x_j)||_{2}^{2}/(2\sigma^2))}{\sum_{j=1,j\neq i}\text{exp}(-||\boldsymbol(x_i)-\boldsymbol(x_j)||_{2}^{2}/(2\sigma^2))} & i \neq j \\
0 & i = j
\end{cases}
\end{equation}
Similarly, we could compute the transition probabilities in the lower-dimension space, which could define:\\
\begin{equation}
q_{j|i} =
\begin{cases}
\frac{\text{exp}(-||\boldsymbol(x^{*}_{i})-\boldsymbol(x^{*}_{j})||_{2}^{2}/(2\sigma^2))}{\sum_{j=1,j\neq i}\text{exp}(-||\boldsymbol(x^{*}_{i})-\boldsymbol(x^{*}_{j})-||_{2}^{2}/(2\sigma^2))} & i \neq j \\
0 & i = j
\end{cases}
\end{equation}
And now our purpose is to fit the model to get the data $\boldsymbol{x^{*}}$ that could preserve the structural information. SNE is a method that consider to minimize the KL divergence between $p_{i|\cdot}$ and $q_{\cdot|i}$, and the target function could be describe as:\\
\begin{equation}\label{SNEtarget}
\begin{split}
l(\boldsymbol(x^{*}),\boldsymbol(x)) &= \sum_{i} KL(p_{i|\cdot}|q_{\cdot|i}) \\
& = \sum_{i}\sum_{j} p_{i|j}log(\frac{p_{i|j}}{q_{j|i}})
\end{split}
\end{equation}
Equation \ref{SNEtarget} could be solved by gradient descent. And t-SNE is a variate of SNE to modify the 'unsymmetrical' KL-divergence. It defines a 'joint' distribution of $p_{i,j}$ with:\\
\begin{equation}
p_{i,j} = \frac{p_{i|j}+p_{j|i}}{2m}
\end{equation}
Where $m$ is the number of samples. The similar method could be used on $q$, and then we would have a 'symmetric' target function.
In this work, we utilized the tSNE package in sk-learn API and specified certain parameters to compute an optimal solution. Notice that in our situation features are partially discrete and partially continuous, thus we compute the distance between samples explicitly with Euclidean for continuous features and matching numbers for discrete features as the following equation:\\
\begin{equation}
\boldsymbol{D_{d}}(\boldsymbol{x_{i}},\boldsymbol{x_{j}}) = \frac{N-N_{m}}{N}
\end{equation}
Where $N$ is the number of features(covariates) $N_{m}$ is the number of features that match in sample $\boldsymbol{x_{i}}$ and $\boldsymbol{x_{j}}$.\\
Computing continuous and discrete features separately could cause another problem: the imbalance of distance. Since the value of $\boldsymbol{D_{d}}(\boldsymbol{x_{i}},\boldsymbol{x_{j}})$ has a maximum value of 1, where the distance of continuous data is not constrained, the effect of $\boldsymbol{D_{d}}(\boldsymbol{x_{i}},\boldsymbol{x_{j}})$ could be under-valued. An approach to deal with this problem is to multiply a correction coefficient $\tau$ to $\boldsymbol{D_{d}}$ with value:\\
\begin{equation}
\tau = \frac{N_{d}}{N_{c}} \times \frac{max(\boldsymbol{D_{c}})}{max(\boldsymbol{D_{d}})}
\end{equation}
Where $N_{d}$ is the number of discrete features and $N_{c}$ is the amount of continuous features. $max(\boldsymbol{D_{c}})$ stands for the maximum value of continuous distance and $max(\boldsymbol{D_{d}})$ demotes the maximum value of discrete features. And the overall distance will be computed by:\\
\begin{equation}\label{tSNEdistanceComputation}
\boldsymbol{D}(\boldsymbol{x_{i}},\boldsymbol{x_{j}}) = \boldsymbol{D_{c}}(\boldsymbol{x_{i}},\boldsymbol{x_{j}}) + \tau * \boldsymbol{D_{d}}(\boldsymbol{x_{i}},\boldsymbol{x_{j}})
\end{equation}
This method is applied in our program, the details will be further illustrated in section 5.
\section{Mathematical Analysis of the Proposed Method}
The novelties of our method mostly lie in the designing of the embedded graph with Gaussian and Delta Kernel (equation \ref{GraphDesignEmbed}) and the clustering-based sampling method (equation \ref{FinalSampleEquation}). The rationality of the new graph-neightbour method is easy to prove, as the kernel methods used are symmetric and positive semi-definite. However, one might have doubt of our proposed method and its assumption and rationality. In this section, we will analyze the assumptions and the mathematical properties of the proposed method, and show that the method could have an analytical upper bound of error rate.\\
The basic assumption has been stated in section 3.2 with equation \ref{BasicAssumption}. Here to further illustrate the assumption, we could denote the sample number of underlying patterns and the classes with the following matrix:\\
\begin{equation}
\begin{bmatrix}
    N_{1}\\
    N_{2} \\
    \cdots        \\
    N_{d}
\end{bmatrix}
=
\begin{bmatrix}
    p_{11}      & p_{12} & \cdots & p_{1n} \\
    p_{21}      & p_{22} & \cdots & p_{2n} \\
    \cdots      & \cdots &        & \cdots      \\
    p_{d1}     &  p_{d2} & \cdots  & p_{dn}
\end{bmatrix}
\begin{bmatrix}
    c_{1}   \\
    c_{2}   \\
    \cdots     \\
    c_{n}
\end{bmatrix}
\end{equation}
Where $N_{i}$ means the sample number of class $i=1,2,...,d$ and $c_{j}$ means the number of samples belonging to underlying pattern $j\in {1,2,...,n}$. The component $p_{ij}$ of the $d\times n$ matrix $\boldsymbol{P}$ denotes the proportion of underlying cluster $j$ to be assigned to category (treatment) $i$. 
According to equation \ref{AssignMatrix}, data after clustering could also be partitioned into different parts with the $d \times n$ matrix $\boldsymbol{W}$. Now we assume that the corrupted data (with mislabeled samples) could be denoted as:\\
\begin{equation}
\begin{bmatrix}
    N^{*}_{1}\\
    N^{*}_{2} \\
    \cdots        \\
    N^{*}_{d}
\end{bmatrix}
=
\begin{bmatrix}
    p^{*}_{11}      & p^{*}_{12} & \cdots & p^{*}_{1n} \\
    p^{*}_{21}      & p^{*}_{22} & \cdots & p^{*}_{2n} \\
    \cdots      & \cdots &        & \cdots      \\
    p^{*}_{d1}     &  p^{*}_{d2} & \cdots  & p^{*}_{dn}
\end{bmatrix}
\begin{bmatrix}
    c_{1}   \\
    c_{2}   \\
    \cdots     \\
    c_{n}
\end{bmatrix}
\end{equation}
And here $\boldsymbol{P^{*}}=p^{*}_{ij}$ could be regarded as the true distribution proportion matrix with corrupted (misclassified/mistreated) samples. And our clustering method is performed upon this matrix. To satisfy our assumptions, the matrix should have the following property:\\
\begin{equation}\label{SummationConstraint}
\sum_{i=1}^{d} p^{*}_{ij} = 1
\end{equation}
To make sure that all samples from each cluster will be fully assigned to different classes. And based on our assumption in equation \ref{MajorityAssumption}, the following property should hold as the number of the samples asymptotically grows to infinity:\\
\begin{equation}\label{InequalityConstraint}
\sum_{i\in \boldsymbol{I}} p^{*}_{ki}c_{i} > \sum_{j \in \boldsymbol{I}} p^{*}_{kj}c_{j}, 	\forall	k \in {1,2,...,d}
\end{equation}
Where $\boldsymbol{I}$ is the set of the pattern symbols that $p_{ki}>0$ and $\boldsymbol{J}$ is the set with originally $p_{kj}=0$, which means that cluster $j$ should not have any member been assigned to class $k$. For each class $k\in {1,2,..,d}$, the whole number of data is composed of:\\
\begin{equation}
N^{*}_{k} = \sum_{i\in \boldsymbol{I}} \frac{p^{*}_{ki}}{\sum_{k=1}^{d}{p^{*}_{ki}}}
+ \sum_{j\in \boldsymbol{J}} \frac{p^{*}_{kj}}{\sum_{k=1}^{d}{p^{*}_{kj}}}
\end{equation}
In the above equation, the first term of the right hand side comes from the clusters that should be assigned to the class and the second term comes as pure mistakes. The error-specified size of sample could then be calculated as:\\
\begin{equation} \label{ErrorDefinition}
\varepsilon = \sum_{i\in \boldsymbol{I}} \frac{|p_{ki}-p^{*}_{ki}|c_{i}}{\sum_{k=1}^{d}{p^{*}_{ki}}}+\sum_{j\in \boldsymbol{J}} \frac{p^{*}_{kj}c_{j}}{\sum_{k=1}^{d}{p^{*}_{kj}}}
\end{equation}
Plugging in the equation \ref{SummationConstraint} and \ref{InequalityConstraint} to equation \ref{ErrorDefinition}, we could get:\\
\begin{equation}\label{ErrorDerivation}
\begin{split}
\varepsilon & = \sum_{i\in \boldsymbol{I}}|p_{ki}-p^{*}_{ki}|c_{i}+\sum_{j\in \boldsymbol{J}} p^{*}_{kj}c_{j} \\
& < \sum_{i\in \boldsymbol{I}}|p_{ki}-p^{*}_{ki}|c_{i}+\sum_{i\in \boldsymbol{I}} p^{*}_{ki}c_{i} \\
& = 
\sum_{i\in \boldsymbol{I_{1}}} p_{ki}c_{i} + 
\sum_{i\in \boldsymbol{I_{2}}} 2(p^{*}_{ki}-p_{ki})c_{i}
\end{split}
\end{equation}
Where $\boldsymbol{I_{1}}$ stands for the set that $p^{*}_{ki} \leqslant p_{ki}$, and $\boldsymbol{I_{2}}$ denotes the set of $i$ that $p^{*}_{ki} \geqslant p_{ki}$. And by computing the error rate $\eta$ with ErrorRate = ErrorSamples/(ErrorSamples + CorrectSamples), we could get the following result from equation \ref{ErrorDefinition}:\\
\begin{equation}\label{ErrorBoundTotal}
\eta < \frac{\sum_{i\in \boldsymbol{I_{1}}}p_{ki}c_{i}+\sum_{i\in \boldsymbol{I_{2}}} 2(p^{*}_{ki}-p_{ki})c_{i}}{\sum_{i\in \boldsymbol{I_{1}}}(p_{ki}c_{i}+p^{*}_{ki}c_{i})+\sum_{i\in \boldsymbol{I_{2}}} 2p^{*}_{ki}c_{i}}
\end{equation}
And now, we could further analyze equation \ref{ErrorBoundTotal} with two special cases.
\begin{enumerate}
\item When $p^{*}_{ki} \leqslant p_{ki}, \text{ }\forall i \in {1,2,...,n}$.\\
Then we will have a number of correct samples as $\sum_{i\in \boldsymbol{I}}p^{*}_{ki}c_{i}$. The error rate would therefore be:\\
\begin{equation} \label{ErrorBoundCaseA}
\eta < \frac{\sum_{i\in \boldsymbol{I}} p_{ki}c_{i}}{\sum_{i\in \boldsymbol{I}}( p_{ki}c_{i} + p^{*}_{ki}c_{i})}
\end{equation}
And equation \ref{ErrorBoundCaseA} would be the upper bound of the error rate under this situation. As we could see, in this case the function is a decrement function w.r.t $p^{*}_{ki}$, and when $p^{*}_{ki} = p_{ki} \forall i$, which means at this special point, the upper bound of error would have a minimal value of $\eta < 0.5$.
\item When $p^{*}_{ki} \geqslant p_{ki} \text{ }\forall i \in {1,2,...,n}$.\\
Then we will have a number of correct samples as $\sum_{i\in \boldsymbol{I}}p_{ki}c_{i}$. The error rate would therefore be:\\
\begin{equation} \label{ErrorBoundCaseB}
\eta < \frac{\sum_{i\in \boldsymbol{I}} 2(p^{*}_{ki}-p_{ki})c_{i}}{\sum_{i\in \boldsymbol{I}} 2p^{*}_{ki}c_{i}}
\end{equation}
The equation \ref{ErrorBoundCaseB} is the upper bound of the second situation and it is an increment function. As we could prove the minimal of this upper bound is also 0.5, and the situation is still $p^{*}_{ki} = p_{ki} \forall i$. This comes with the property of a continuous upper bound of the method.
\end{enumerate}
And in the above method we have shown the assumptions of the method and why it is rational. In practice, if the clustering algorithm is strong enough to capture the majority of correct-specified samples, then $p^{*}_{ki}$ should be close to $p_{ki}$ given the assumption in \ref{MajorityAssumption}.
\section{Practical Settings and Experimental Results}
In addition to algorithm designing and analyzing, in this work parameter setting problems are also discussed and experimental results are obtained based on SEER data. This section will first introduce the SEER data, and then discuss (hyper)parameter setting problems. And finally, the experimental results with our method will be demonstrated.
\subsection{SEER Dataset}
The SEER dataset from 2004 to 2014 was queried for patients who were confirmed pancreatic adenocarcinoma defined according to the International Classification of Disease for Oncology, Third Edition (ICD-O-3) codes for morphology (8140 and 8500) and topography (C25.0, C25.1, C25.2, C25.3, C25.7, C25.8, and C25.9). Data collected for each patient includes patient characteristics (age at diagnosis, sex, race, and marital status), tumor characteristics (tumor location, tumor grade, and AJCC stage), and treatment characteristics (type of surgery, radiotherapy and status of surgery adjuvant radiotherapy). These three treatments are considered as the three classes/categories under the term of Machine Learning in out work.\\
There are some missing data in the dataset, and discussing them will be out of our research scope. Thus, in our program we only consider the features(covariates) without data missing, and the rationality of the selection of features could be supported from the perspective of bioscience and medical science. According to the true meaning of each feature, there only four features which could be regarded as continuous: EOD10\_PN, CS\_SIZE, CS\_EXT, and CS\_NODE. Some of the features could imply the treatment (say, NO\_SURG flag) or should not be treated as feature (say, CASENUM, the index of patient), so we also remove these features.\\
The original data is believed to only have mild mis-labeled samples (mis-treated patients). To demonstrate the performance of our method under data with different extent of corruption, we also generate some further-corrupted data for training purpose. We produce data with 10\%, 20\% and 40\% of artificial label corruptions based on the original data by corrupting the labels to other two classes with equal opportunities. Notice that the artificially corrupted data is only utilized for training, and when it comes to evaluation, we will use the original data to test the robustness of our proposed approach.
\subsection{Parameter Settings}
Following the method proposed in section 3, there are some hyper- parameters that should be assigned based on our assumption and practical Requirement. Here we consider three aspects of parameter setting, namely the number of clusters, Spectral Clustering and Xgboost parameter settings, and considerations of tSNE.\\
The first hyper-parameter to determine is the number of clusters we would like to use. This reflects the our basic assumption of the number of underlying patterns of the data. In our program we choose $n=d$ where $n$ is the number of clusters and $d$ as the number of categories (classes/treatments). This setting is for a better interpretation property: if there are no overlaps between each classes (mutually exclusive, say partial gastrectomy and total gastrectomy), then ideally the clustering result should have a distinct distribution between different classes and the correctly labeled samples will be straightforward to discover; otherwise, if there are overlaps between different classes/treatments(say in our SEER data situation, there are operation treatment, radian treatment and operation+radian treatment), then this setting will be convenient for us to analyze the demanded components of each class. In the subsection 5.3 we will demonstrate the interpretation of the SEER data we used in the experiment based on this scheme of cluster number setting.\\
The second topic of parameter setting is our selections for the Spectral Clustering and Xgboost API. In our work, we use the Spectral Cluster method provided by SK-learn package with Python API. In the package, we specify to use Arpack to compute the eigenvalue decomposition. This will lead to a slightly higher time consumption, but the result will be significantly more stable. The final clustering results are obtained by K-mean over the embedded space, and to obtain a close-to-optimal result, the method will compute K-means in the embedded space with 10 different initializations to get the best result. And in our program, the $\sigma$ value used in equation \ref{GaussianKernel} could either be specified or be computed via equation \ref{SigmaComputing}. The parameters of Xgboost has been discussed in section 3.3. In practice, we set tree maximum depth as 7, and $\eta$ value as 0.5 to control the learning with a relatively conservative scheme. We specified the max training iteration as 10 rounds for original data and 5 rounds for artificially corrupted data, as this setting could provide as an optimal validation error.\\
The third point to be considered in practice is the tSNE visualization technique. In this project, we use the tSNE package provided by SK-learn. The authors of the original paper proposed tSNE suggested that hyper-parameters of tSNE is robust and won't make too large an effect on the result. However, in our practical experiment, things are much more tricky: the parameter of 'perplexity', which determines the balance between local manifold and global manifold, could significantly affect the clustering result and the optimal setting of this parameter will vary with the number of samples for visualization. Briefly speaking, the quantity of perplexity should be in the same magnitude of number of samples. For example, if we have 300 samples, then a perplexity of 5-50 will be good for visualization; however, if we have 30000 sample, then it seems the value of perplexity score should be adjusted to a larger value accordingly. \\
In our program, we visualized the project of data in different clusters and treatments(classes) with tSNE respectively. For the cluster illustration, we randomly choose 50 samples from each cluster (150 sample in total) and set perplexity=90, learning\_rate = 7 and use our precomputed distance with equation \ref{tSNEdistanceComputation}; And for the classes illustration problem, we choose 100 samples from each treatment group (before and after processed by our sampling method) and set perplexity=180 and learning\_rate = 7. 
\subsection{Experimental Result}
\begin{figure}[h]
\centering\includegraphics[width=0.5\linewidth]{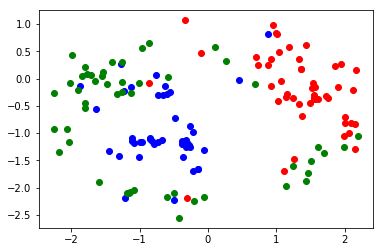}
\caption{Clustering Result in 2d. In the figure, \textbf{red samples are from cluster 1, blue samples are from cluster 2 and green samples are from cluster 3}. The group of red samples has a clear boundary with the rest two clusters, comparing the relatively vague boundary between cluster 2 and 3}
\label{tSNEclusteringResult}
\end{figure}

\begin{figure}[p]
\centering\includegraphics[width=0.5\linewidth]{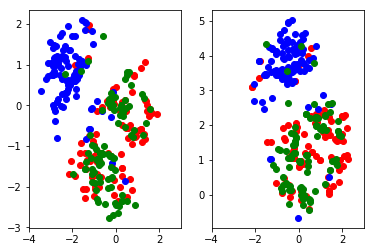}
\caption{The comparison of data projection in 2-d plane of training set before and after the proposed process. left: data manifold before process; right: data manifold before process. In the figure, \textbf{red samples are from treatment 1, blue samples are from treatment 3 and green samples are from treatment 3}. }
\label{tSNEdataSeparation}
\end{figure}

\begin{figure}[p]
\centering\includegraphics[width=0.5\linewidth]{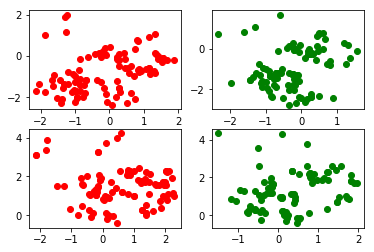}
\caption{The comparison of data projection inside treatment 1 and 2 in 2-d plane of training set before and after the proposed process. In the figure, \textbf{red samples are from treatment 1 and green samples are from treatment 3}. top: data manifold before process; bottom: data manifold before process. From the figure, we could find out that data manifold after our process tends to have some difference between treatment 1 and 3, although they are still very similar.}
\label{tSNEtreatment1and3}
\end{figure}

There are in total 8683 valid records, with 3531 being treated with method 1, 3200 with method 2 and 1952 with method 3. And the result of clustering divide the data into a proportion of 2905 in cluster 1, 3817 in cluster 2 and 1961 in cluster 3. Specifically, the treatment-cluster data segment result could be shown in table \ref{ClusteringResultTable}:\\
\begin{table}[h]
\centering
\begin{tabular}{| l | l | l | l|}
\hline
 & Cluster 1 & Cluster 2 & Cluster 3 \\
\hline
New Treatment 1 & 210 & 2159 & 1162\\
\hline
New Treatment 2 & 2615 & 250 & 335 \\
\hline
New Treatment 3 & 80 & 1408 & 464 \\
\hline
\end{tabular}
\caption{Treatment-Clustering Result}
\label{ClusteringResultTable}
\end{table}
And the sampling probability computed by equation \ref{FinalSampleEquation} could be illustrated in table \ref{SampleProbabilityTable}.\\
\begin{table}[h]
\centering
\begin{tabular}{| l | l | l | l|}
\hline
 & Cluster 1 & Cluster 2 & Cluster 3 \\
\hline
Treatment 1 & 0.0587 &  0.4597 & 0.4816\\
\hline
Treatment 2 & 0.7921 &  0.0576 &  0.1503 \\
\hline
Treatment 3 & 0.0435 &  0.5827 &  0.3738 \\
\hline
\end{tabular}
\caption{Probability to sample for new data}
\label{SampleProbabilityTable}
\end{table}
As we mentioned in the previous sections, the setting of $d=n$ could helps us explain the result. In table \ref{ClusteringResultTable} and \ref{SampleProbabilityTable}, it could be derived that: 1. patient characterized by cluster 1 should be \textbf{predominately treated with treatment 2 (radian therapy)}; 2. patients characterized by cluster 3 and cluster 1 prefer to be treated by a treatment \textbf{involved with operations}, but those who belongs to cluster 3 are more likely to \textbf{reject radian therapy}.\\
The illustration of the clustering result in 2-dimension could be shown as figure \ref{tSNEclusteringResult}. The result is produced with the settings stated in section 5.2 and we could find out that the three clusters have clear boundaries against each other, implying that the result should be rational. The red points, which represent data from cluster 1, have a more distinct boundary comparing to cluster 2 and 3, and this could be a supporting evidence for patients characterized by cluster 1 to be treated by a separate method.\\
The 2-d projection of training sets before and after process could be illustrated in figure \ref{tSNEdataSeparation}. The figure could prove the rationality of our process and the overlap between treatment 1 and 3. Meanwhile, we could find out the data manifolds are more compact and have multiple 'clusters' inside the class, indicating that the dataset after process could be better categorized comparing to those before process. This is more significant in figure \ref{tSNEtreatment1and3} as we specifically illustrate the manifold of treatment 1 and treatment 3.
\begin{figure}
\begin{subfigure}[h]{1.6in}
\centering\includegraphics[width=\linewidth]{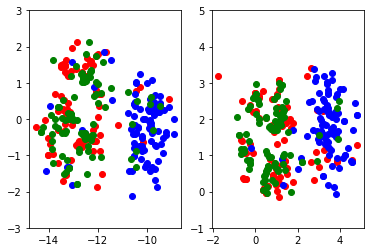}
\caption{Comparison of data with 10\% corrupted}
\end{subfigure}
\begin{subfigure}[h]{1.6in}
\centering\includegraphics[width=\linewidth]{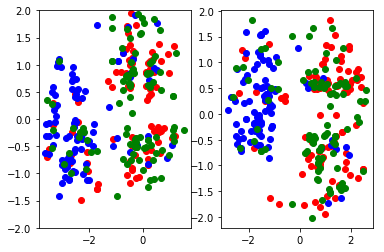}
\caption{Comparison of data with 20\% corrupted}
\end{subfigure}
\begin{subfigure}[h]{1.6in}
\centering\includegraphics[width=\linewidth]{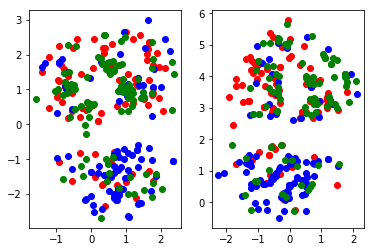}
\caption{Comparison of data with 40\% corrupted}
\end{subfigure}
\caption{Data manifold before and after sampling procedure for data that added artificial label corruptions. \textbf{red samples are from treatment 1, blue samples are from treatment 3 and green samples are from treatment 3}. For each sub-figure, left: before precess, right: after process}
\label{ComparisonDataCorrupted}
\end{figure}
\begin{table}[h]
\centering
\begin{tabular}{| p{3.2cm} | p{1cm}| p{1cm} | p{1.5cm} | p{1.5cm} | p{1.5cm} | p{1.5cm} |}
\hline
 & Age & Race & CS NODE & CS SIZE &  D AJCC S & Survive Month\\
\hline
Processed Data + Xgboost & 0.165 &  0.0025 & 0.0085 & 0.2054 & 0.0695 & 0.3118\\
\hline
Original Data + Xgboost & 0.138 &  0.0241 &  0.1096 & 0.0519 & 0.1377 & 0.1315\\
\hline
Raw Data without Propensity Weighting& 0.089 &  0.0446 &  0.2346 & 0.3857 & 0.7434 & 0.5806\\
\hline
\end{tabular}
\caption{Standardized Bias between treatment group 1 and 2}
\label{TreatmentGroup12Comparison}
\end{table}
\begin{table}[h]
\centering
\begin{tabular}{| p{3.2cm} | p{1cm}| p{1cm} | p{1.5cm} | p{1.5cm} | p{1.5cm} | p{1.5cm} |}
\hline
 & Age & Race & CS NODE & CS SIZE &  D AJCC S & Survive Month\\
\hline
Processed Data + Xgboost & 0.052 &  0.0085 & 0.1597 & 0.2363 & 0.1560 & 0.0218\\
\hline
Original Data + Xgboost & 0.134 &  0.0051 &  0.3821 & 0.4261 & 0.3287 & 0.0714\\
\hline
Raw Data without Propensity Weighting& 0.137 &  0.0605 &  0.0364 & 0.0875 & 0.0307 & 0.1113\\
\hline
\end{tabular}
\caption{Standardized Bias between treatment group 1 and 3}
\label{TreatmentGroup13Comparison}
\end{table}
\begin{table}[h]
\centering
\begin{tabular}{|p{3.2cm}  | p{1cm}| p{1cm} | p{1.5cm} | p{1.5cm} | p{1.5cm} | p{1.5cm} |}
\hline
 & Age & Race & CS NODE & CS SIZE &  D AJCC S & Survive Month\\
\hline
Processed Data + Xgboost & 0.238 &  0.0134 & 0.0981 & 0.3440 & 0.2042 & 0.3237\\
\hline
Original Data + Xgboost & 0.297 &  0.0357 &  0.1130 & 0.3157 & 0.4209 & 0.2432\\
\hline
Raw Data without Propensity Weighting& 0.246 &  0.0195 &  0.1997 & 0.4235 & 0.8343 & 0.8037\\
\hline
\end{tabular}
\caption{Standardized Bias between treatment group 2 and 3}
\label{TreatmentGroup23Comparison}
\end{table}
\begin{table}[h]
\centering
\begin{tabular}{| p{3.2cm} | p{1cm}| p{1cm} | p{1.5cm} | p{1.5cm} | p{1.5cm} | p{1.5cm} |}
\hline
 & Age & Race & CS NODE & CS SIZE &  D AJCC S & Survive Month\\
\hline
Processed Data + Xgboost & 0.053 & 0.0269 & 0.0932 & 0.3351 & 0.4977 & 0.3784\\
\hline
Original Data + Xgboost & 0.045 &  0.0666 &  0.0811 & 0.1496 & 0.4988 & 0.4201\\
\hline
\end{tabular}
\caption{Standardized Bias between treatment group 1 and 2 with 10\% labels corrupted}
\label{TreatmentGroup12Comparison10corrupted}
\end{table}
\begin{table}[h]
\centering
\begin{tabular}{| p{3.2cm} | p{1cm}| p{1cm} | p{1.5cm} | p{1.5cm} | p{1.5cm} | p{1.5cm} |}
\hline
 & Age & Race & CS NODE & CS SIZE &  D AJCC S & Survive Month\\
\hline
Processed Data + Xgboost & 0.059 & 0.0553 & 0.0149 & 0.2696 & 0.4712 & 0.4745\\
\hline
Original Data + Xgboost & 0.037 &  0.0474 &  0.1238 & 0.2757 & 0.5243 & 0.5577\\
\hline
\end{tabular}
\caption{Standardized Bias between treatment group 1 and 2 with 20\% labels corrupted}
\label{TreatmentGroup12Comparison20corrupted}
\end{table}
\begin{table}[h]
\centering
\begin{tabular}{| p{3.2cm} | p{1cm}| p{1cm} | p{1.5cm} | p{1.5cm} | p{1.5cm} | p{1.5cm} |}
\hline
 & Age & Race & CS NODE & CS SIZE &  D AJCC S & Survive Month\\
\hline
Processed Data + Xgboost & 0.0250 & 0.0197 & 0.0968 & 0.3667 & 0.5715 & 0.4491\\
\hline
Original Data + Xgboost & 0.0968 &  0.0244 &  0.1230 & 0.3194 & 0.5793 & 0.5826\\
\hline
\end{tabular}
\caption{Standardized Bias between treatment group 1 and 2 with 40\% labels corrupted}
\label{TreatmentGroup12Comparison40corrupted}
\end{table}
To evaluate the quality of the computed propensity score, we need to compute the confounding metric after applying propensity score to the original data. Here we consider propensity score weighting and employ \textbf{Standardized Bias (SB)} for several key features under ATE settings according to \cite{PropensityScoreEvaluationTutorial}. The metric could denote the level of confound between two groups (treatment and control groups) for each feature (covariate) and if the confounding level is lower (smaller in the metric) then it indicate that the data after propensity score weighted is better randomized. The evaluation method could only compare two groups (classes) so that here we have ${3\choose 2} = 3$ combinations. The comparison of the Standardized Bias could be found in table \ref{TreatmentGroup12Comparison}, \ref{TreatmentGroup13Comparison} and \ref{TreatmentGroup23Comparison}. Here, the Processed Data was trained with xgboost for 10 iterations and the Original Data is a random subset of 3000 samples from the whole dataset. Final training and testing mlogloss on Processed Data are 0.243960 and 0.552677; Final training and testing mlogloss on Original Data are 0.222134 and 0.500497.\\
To demonstrate the effect of noise-correction for further corrupted data, we also compared the Standardized Bias between treatment 1 and treatment 2 for the data with 10\%, 20\% and 40\% artificial corrupted label in table \ref{TreatmentGroup12Comparison10corrupted}, \ref{TreatmentGroup12Comparison20corrupted} and \ref{TreatmentGroup12Comparison40corrupted} respectively. Notice here when computing the Standardized Bias with the whole dataset, we use the authentic data. This will be the common setting in real-life situations. Also notice that for the corrupted experiments the xgboost only computed for 5 iteration for the overfitting consideration.\\
From the table, we could find that 1. propensity score computed by processed data with xgboost could usually outperform the scores computed by original sampled data with the same method; 2. the advantages of processed data will increase as the proportion of corrupted data increase 3. xgboost generally has a excellent and robust performance in computing propensity scores, and the propensity score computed could still improve the randomization even with 40\% corrupted labels.
\section{Conclusion}
In this paper, a novel approach to deal with the corrupted-label problem in computing propensity score for medical data is proposed and its property is analyzed. The Xgboost approach is utilized to compute the propensity score and the performance of the integrated approach is examined and compared. TSNE method is employed for comprehensive purpose and the data manifold of different clusters and treatments are demonstrated.\\
The paper made following major contributions to the field: Firstly, the paper finds out the mislabeling-data problem in medical science and researched on this topic. Given the fact that the problem has been under-investigated before but the phenomenon is very common, the research would significantly fill the blank of this aspect in bioscience research; Secondly, the paper proposed an effective method for getting a relatively authentic subset of training data, and proved mathematically about its rationality and effectiveness. The rigorous proof makes it possible for this method to be utilized into general Machine Learning problems and solve a broader scope of tasks. Thirdly, the paper employed Xgboost and tSNE methods to this area, which are relatively novel attempts. The results imply a huge research potential in this area.\\
In the future, the author would like to concentrate more on mislabeling data in bioscience area. We intend to try penalization-based label correction methods proposed in the field of Machine Learning, and design new algorithms specifically in dealing with medical data. Meanwhile, to analyze the confounding covariates, we plan to research on graphical model-based methods in the long run.

\clearpage



\bibliographystyle{model1-num-names}
\bibliography{sample.bib}






\end{document}